\documentclass[twocolumn]{pasj02}

\RequirePackage{mathpazo}
\RequirePackage[T1]{fontenc}

\def\commenta{$^*$}
\def\commentb{$^\dagger$}
\def\commentc{$^\ddagger$}
\def\commentd{$^\S$}
\def\commente{$^\|$}

\def\earlysh{0.05711(4)~} 
\def\stageAsh{0.05919(5)~} 
\def\stageBsh{0.058190(5)~} 
\def\stageCsh{0.05791(2)~} 
\def\pdot{11.2(3)$\times 10^{-5}$~}
\def\massratio{0.098(4)~}

\def\stageBshV{0.05810(1)~} 
\def\lateshV{0.05811(2)~} 
\def\pdotV{9.6(1.3)$\times 10^{-5}$~}

\newcounter{author}
\def\authorcount#1#2{\refstepcounter{author}\label{#1}
                     \altaffiltext{\ref{#1}}{#2}}

\Received{$\langle$reception date$\rangle$}
\Accepted{$\langle$acception date$\rangle$}
\Published{$\langle$publication date$\rangle$}

\begin{document}
\SetRunningHead{Y. Tampo et al.}{WZ Sge-type dwarf nova ASASSN-24hd}

\title{ASASSN-24hd; a dwarf nova bridging WZ Sge-type and SU UMa-type superoutbursts}

\author{
    Yusuke~\textsc{Tampo}\altaffilmark{\ref{affil:saao}*}$^,$\altaffilmark{\ref{affil:uct}},
    Naoto~\textsc{Kojiguchi}\altaffilmark{\ref{affil:KyotoOkayama}}$^,$\altaffilmark{\ref{affil:Kyoto}},
    Taichi~\textsc{Kato}\altaffilmark{\ref{affil:Kyoto}},
    Mariko~\textsc{Kimura}\altaffilmark{\ref{affil:kanazawa}}, 
    David. A. H.~\textsc{Buckley}\altaffilmark{\ref{affil:saao}}$^,$\altaffilmark{\ref{affil:uct}}$^,$\altaffilmark{\ref{affil:ufs}},
    Berto~\textsc{Monard}\altaffilmark{\ref{affil:mlf1}}$^,$\altaffilmark{\ref{affil:mlf2}}, 
    Franz-Josef~\textsc{Hambsch}\altaffilmark{\ref{affil:ham1}}$^,$\altaffilmark{\ref{affil:ham2}}$^,$\altaffilmark{\ref{affil:dfs}}, 
    Katsuki~\textsc{Muraoka}\altaffilmark{\ref{affil:Kyoto}},
    Daisaku~\textsc{Nogami}\altaffilmark{\ref{affil:Kyoto}},
    Stephen B. ~\textsc{Potter}\altaffilmark{\ref{affil:saao}}$^,$\altaffilmark{\ref{affil:ujoh}},
    Anke~\textsc{van~Dyk}\altaffilmark{\ref{affil:saao}}$^,$\altaffilmark{\ref{affil:uct}},
    and Patrick~\textsc{Woudt}\altaffilmark{\ref{affil:uct}} 
}

\authorcount{affil:saao}{
    South African Astronomical Observatory, PO Box 9, Observatory, 7935, Cape Town, South Africa}
\email{$^*$yusuke@saao.ac.za}

\authorcount{affil:uct}{
    Department of Astronomy, University of Cape Town, Private Bag X3, Rondebosch 7701, South Africa}

\authorcount{affil:KyotoOkayama}{
     Okayama Observatory, Kyoto University, 3037-5 Honjo, Kamogatacho,
     Asakuchi, Okayama 719-0232, Japan}
     
\authorcount{affil:Kyoto}{
     Department of Astronomy, Kyoto University, Kitashirakawa-Oiwake-cho, Sakyo-ku, 
     Kyoto 606-8502, Japan}

\authorcount{affil:kanazawa}{
    Advanced Research Center for Space Science and Technology, Colledge of Science and 
    Engineering, Kanazawa University, Kakuma, Kanazawa, Ishikawa 920-1192}

\authorcount{affil:ufs}{
    Department of Physics, University of the Free State, P.O. Box 339, Bloemfontein 9300, South Africa}

\authorcount{affil:mlf1}{
    Bronberg Observatory, Center for Backyard Astrophysics Pretoria, PO Box 11426, Tiegerpoort 0056, South Africa}

\authorcount{affil:mlf2}{
    Kleinkaroo Observatory, Center for Backyard Astrophysics Kleinkaroo, Sint Helena 1B, PO Box 281, Calitzdorp 6660, South Africa}

\authorcount{affil:ham1}{
    Groupe Européen d’Observations Stellaires (GEOS), 23 Parc de Levesville, 28300 Bailleau l’Evêque, France}
    
\authorcount{affil:ham2}{
    Bundesdeutsche Arbeitsgemeinschaft für Veränderliche Sterne (BAV), Munsterdamm 90, 12169 Berlin, Germany}

\authorcount{affil:dfs}{
    Vereniging Voor Sterrenkunde (VVS), Oostmeers 122 C, 8000 Brugge, Belgium}

\authorcount{affil:ujoh}{
    Department of Physics, University of Johannesburg, PO Box 524, Auckland Park 2006, South Africa}


\KeyWords{accretion, accretion disk --- novae, cataclysmic variables ---- stars: dwarf novae
 --- stars :individual (ASASSN-24hd)}

\maketitle

\begin{abstract}

WZ Sge-type dwarf novae (DNe) form a subclass in cataclysmic variables, characterized by short-period variations called superhumps during an outburst.
Here we present optical ground-based and TESS observations of ASASSN-24hd in its 2024-2025 outburst. ASASSN-24hd is the first reported WZ Sge-type DN outburst fully covered by TESS, providing a great opportunity to study the evolution of superhumps. Our observations establish its early and stage-A ordinary superhumps as \earlysh and \stageAsh d, respectively, resulting in its mass ratio of \massratio. The TESS observations confirm that the evolution of its superhump period, amplitude, and profile after the appearance of ordinary superhumps is generally consistent with those of SU UMa-type DNe observed with Kepler and TESS.
Furthermore, we find that ASASSN-24hd in outburst shares a great similarity to the 2010 superoutburst of an SU UMa-type DN V585 Lyr, observed by Kepler, particularly including the superhump evolution and the long waiting time ($\gtrsimeq$ 5 d) before the stage A--B transition of ordinary superhumps. 
The shorter superoutburst cycles and smaller outburst amplitude in V585 Lyr than those of ASASSN-24hd disfavor the interpretation that V585 Lyr is, in fact, a face-on WZ Sge-type DN where early superhumps are undetectable.
Instead, one possibility of their critical differences is either low quiescence viscosity or inner disk truncation, which has been invoked to explain the extreme nature of WZ Sge-type DNe, but future observations in quiescence are vital to conclude. These findings emphasize the borderline between SU UMa-type and WZ Sge-type DNe.

\end{abstract}

\section{Introduction}
\label{sec:1}

Cataclysmic variables (CVs) are close binary systems that host an accreting white dwarf (WD) and a mass-transferring low-mass secondary star that fills its Roche lobe. An accretion disk is formed around the WD if the WD is not strongly magnetized (see \citet{war95book,hel01book} for a general review). Dwarf novae (DNe) are a subclass of non-magnetic CVs, showing recurrent outbursts in an accretion disk. The mechanism of DN outbursts is understood in terms of the thermal instability model in a disk \citep{osa96review, kim20thesis, ham20CVreview}, in which a rapid increase in disk viscosity along with hydrogen ionization enhances the disk accretion rate, and the released gravitational energy is observed as an outburst.

Along with short-lived normal outbursts, DNe with a mass ratio $q$ lower than $\approx 0.3$ exhibit a large-amplitude and long-duration outburst called a superoutburst accompanying ordinary superhumps in the light curve, a saw-shaped single-peaked variation with a period slightly (2--3 \%) longer than the orbital period $P_{\rm orb}$ (see \citet{Pdot} and their following series for various examples). 
DNe showing superoutbursts are classified as an SU UMa-type DN. Ordinary superhumps exhibit three stages based on their period evolution \citep{Pdot}.  The growing superhumps with the constant and longest superhump period $P_{\rm sh}$ are called stage-A superhumps. Stage-B superhumps show decreasing amplitude and increasing period. Period derivatives ($P_{\rm dot} = \dot{P}_{\rm sh} / P_{\rm sh}$) are usually defined for stage-B superhumps. Stage-C superhumps follow with shorter periods. In the tidal-thermal instability (TTI) model \citep{whi88tidal, osa89suuma, hir90SHexcess, osa13v1504cygKepler}, the growth of tidal instability at the 3:1 resonance radius corresponds to stage-A superhumps. The start of the inward propagation of the eccentricity alters the system to stage-B superhumps. The short superhump period during the stage-C superhumps can be associated with the smaller disk radius at the end of a superoutburst. In SU UMa-type DNe with long ($\geq$ 2 years) superoutburst cycles, a halt of outburst decline is observed $\simeq$ 5 d before the start of rapid decline \citep{bab00v1028cyg, kat03hodel}. This halt often accompanies the stage B--C superhump transition and regrowth of superhump amplitude.

WZ Sge-type DNe form a subclass in SU UMa-type DNe, characterized by the shortest orbital period, largest outburst amplitude, longest outburst cycle (a decade or longer), and lack of normal outbursts (\citet{kat15wzsge} for a review). These characteristics have been considered to be attributed to either (1) very low disk viscosity in quiescence due to lower mass-transfer rate \citep{sma93wzsge, osa95wzsge} or (2) the inner disk truncation by the WD magnetosphere or the evaporation effect \citep{ham97wzsgemodel, mey98wzsge, mat07wzsgepropeller}. 
The modern criterion for WZ Sge-type DNe in outburst is the detection of double-peaked modulation, called early superhumps, observed during the first 5--10 days of the superoutburst and before the appearance of ordinary superhumps. The period of early superhumps is approximately equal ($\approx 0.1\%$ level) to the orbital one \citep{ish02wzsgeletter}. Since the amplitude of early superhumps shows a dependence on the inclination of the systems \citep{uem12ESHrecon, kat22WZSgecandle}, a long waiting time ($\gtrsimeq$ 5 d) before the appearance of ordinary superhump is also regarded as a characteristic of low-inclination WZ Sge-type DNe.

\citet{osa02wzsgehump} propose that the 2:1 resonance excited in a disk \citep{lin79lowqdisk, kun05earySHSPH} is the origin of early superhumps, which naturally explains its double-peaked profile, inclination-dependent amplitude, and why the early superhumps are observed only in low-mass ratio (typically $q~\lesssim$ 0.1) systems.
However, recent studies using high-quality and continuous observations by Kepler and TESS have reported modulations resembling early superhumps at the beginning of superoutbursts in K2BS5 ($P_{\rm orb}$ = 0.08580(6) d; \cite{boy24k2bs5}) and V844 Her ($P_{\rm orb}$ = 0.054643 d; \cite{kat22v844her}). These modulations show a double-peaked profile and period close to the orbital one (i.e. definitely shorter than that of the following ordinary superhumps), although they last only for 1--2 d compared to 5--10 d in genuine early superhumps. They concluded that these pre-stage-A oscillations are not early superhumps because of (1) the short duration of pre-stage-A oscillations on K2BS5, and (2)  the average epoch of humps in this phase being located on the smooth extension of stage-A superhumps on V844 Her.
BC UMa ($P_{\rm orb}$ = 0.06261(4) d) shows similar double-wave variations with the orbital period for the first four and two days in its 2000 \citep{pat03suumas} and 2003 \citep{mae07bcuma} superoutbursts, respectively. Given the large mass ratio at that time ($q=0.13$ but now refined to $q=$0.096(6) in \citet{wak21a18aan}), \citet{mae07bcuma} interpreted this variation as the tidally-induced spiral shocks on the accretion disk at the tidal truncation radius.
V585 Lyr is another puzzling SU UMa-type DN, with the 0.06041(1)-d superhump period \citep{kat13j1939v585lyrv516lyr}. This system shows normal outbursts and superoutbursts with a precursor outburst with $\simeq$ 2.3-yr superoutburst cycle, typical outburst behavior of an SU UMa-type DN \citep{kry01v585lyrv587lyr, Pdot8}. 
However, its 2010 superoutburst lacked a precursor outburst, and the waiting time before the stage A--B superhump transition took $\simeq$ 7 d \citep{kat13j1939v585lyrv516lyr}. \citet{kat13j1939v585lyrv516lyr} interpreted this as the type B superoutburst in \citet{osa03DNoutburst, osa05DImodel}; the mass stored at the onset of the superoutburst is large enough and the disk can remain at or beyond the 3:1 resonance radius for some time before the 3:1 resonance and superhumps start to grow.
Meanwhile, there is an increasing population in WZ Sge-type DNe which has shown both types of superoutbursts; longer and brighter ones accompanying early superhumps, but also shorter and fainter ones lacking early superhump phase (e.g. the 2015 superoutburst of AL Com, rebrightening superoutbursts in V3101 Cyg; \cite{kim16alcom, tam20j2104}). WZ Sge-type DNe with a mass ratio larger than 0.1 have also been found  (\citet{wak17asassn16eg} and references therein). These authors discuss that the mass ratio determines only if the disk is able to excite the 2:1 resonance but individual superoutbursts with early superhump phase are achieved when the disk mass at the onset outburst is large enough for the disk to reach the 2:1 resonance radius. 
Thus a study of CVs at the border of SU UMa-type and WZ Sge-type DNe can provide a view of how these two types of superoutbursts are connected and can be distinguished from each other.

Here we report the detailed observations of ASASSN-24hd during its outburst in 2024, including the coincident TESS observation covering this entire outburst, the first reported case in WZ Sge-type DNe.
ASASSN-24hd was initially discovered by ASAS-SN \citep{ASASSN} as a CV candidate on 2024-12-17.03 UTC. 
Time-resolved observations suggested that ASASSN-24hd is a WZ Sge-type DN through the development of superhumps $\approx$ 6 d after the outburst detection (vsnet-alert 28039\footnote{<http://ooruri.kusastro.kyoto-u.ac.jp/mailarchive/vsnet-alert/28039>}).
The quiescence counterpart is Gaia EDR3 4641960005946996096 with $G = 19.151(4)$ mag at 318.9$^{+18.6}_{-17.8}$ pc, corresponding to the quiescence absolute magnitude $M_{G \rm - min} =  11.6(1) $ mag without considering Galactic extinction \citep{gaiaedr3, Bai21GaiaEDR3distance}.

This paper is structured as follows;  section \ref{sec:2} summarizes an overview of our ground-based and TESS observations. Section \ref{sec:3} presents the result of the photometric analysis of the overall superoutburst and superhumps of ASASSN-24hd. The result of our spectroscopic observations is shown in section \ref{sec:4}. We discuss the classification of ASASSN-24hd as a WZ Sge-type DN, its comparison with V585 Lyr, and superhump evolution in section \ref{sec:5}. All the observation epochs in this paper are described in the Barycentric Julian Day (BJD).

\section{Observations and Analysis}
\label{sec:2}

\subsection{ground-based time-resolved observations}
Our ground-based time-resolved CCD photometric observations of ASASSN-24hd were carried out through the Variable Star Network (VSNET) collaboration \citep{VSNET}. We also obtained the time-series observations using Mookodi \citep{era24mookodi} on the Lesedi telescope and the Simultaneous-Color InfraRed Imager for Unbiased Survey (SIRIUS; \cite{nag99sirius, tag03sirius}) on the IRSF 1.4m telescope, both located at the Sutherland Observatory of the South African Astronomical Observatory. Their instrument details and observation logs are summarized in tables E1 and E2 \footnote{Tables E1--E4 are available only on the online edition as Supporting Information.}, respectively. The zero point of the observations unfiltered or with the clear band was adjusted to the $V$ band of standard stars ($CV$ band).
Before period analysis, the global trend of the light curve was removed by subtracting a smoothed light curve obtained by locally weighted polynomial regression (LOWESS: \cite{LOWESS}) with a typical timescale of $\simeq$ 0.5--1.0 d in outburst. The superhump maxima were determined following \citet{Pdot, kat13j1939v585lyrv516lyr}. The phase dispersion minimization (PDM;  \cite{PDM}) method was applied for a period analysis of superhumps in this paper.  The 1$\sigma$ error for the PDM analysis is determined following \citet{fer89error, pdot2}.

\subsection{archival time-domain surveys}

We also extracted the all-sky photometric survey data from the All-Sky Automated Survey for SuperNovae (ASAS-SN) Sky Patrol \citep{ASASSN, koc17ASASSNLC, ASASSNV2}, the Asteroid Terrestrial-impact Last Alert System (ATLAS; \cite{ATLAS, hei18atlas, smi20atlas, shi21atlas}), and the Gaia Alerts (Gaia25acw\footnote{<http://gsaweb.ast.cam.ac.uk/alerts/alert/Gaia25acw/>}) to examine the global light curve profile in and before the outburst.
Lastly, ASASSN-24hd was in the field of view of the Transiting Exoplanet Survey Satellite (TESS) sector 87, Camera 4, CCD 3. We downloaded the TESS Image CAlibrator Full Frame Images (TICA FFIs; \cite{fau20tica}) via the MAST portal\footnote{<https://archive.stsci.edu/hlsp/tica>; DOI 10.17909/t9-9j8c-7d30} and performed aperture photometry by ourselves using the \texttt{lightkurve} package \citep{lightkurve}. We describe the details of our photometry procedure of TESS in the appendix. No objects are found in the previous TESS sectors at the position of ASASSN-24hd, likely because it was too faint for the TESS pipelines.

\subsection{Spectroscopic observations}

We performed optical spectroscopic observations using the Spectrograph Upgrade: Newly Improved Cassegrain (SpUpNIC; \cite{spupnic}) mounted on the 1.9-m telescope at the Sutherland Observatory of the South African Astronomical Observatory on BJD 2460671.34 and 2460674.29, with the total exposure time of 1200 and 900 sec, respectively.
We used the reflection grating 6 with a resolution of $R\sim 3500$ and wavelength coverage of 4415--7050\AA. Data reduction was performed using IRAF in the standard manner (bias subtraction, flat fielding, aperture determination, spectral extraction, wavelength calibration with arc lamps, and flux normalization by continuum).

\section{Results}
\label{sec:3}

\subsection{Overall light curve of outburst}
\label{sec:3.1}

\begin{figure*}[tbp]
 \begin{center}
  \includegraphics[width=\linewidth]{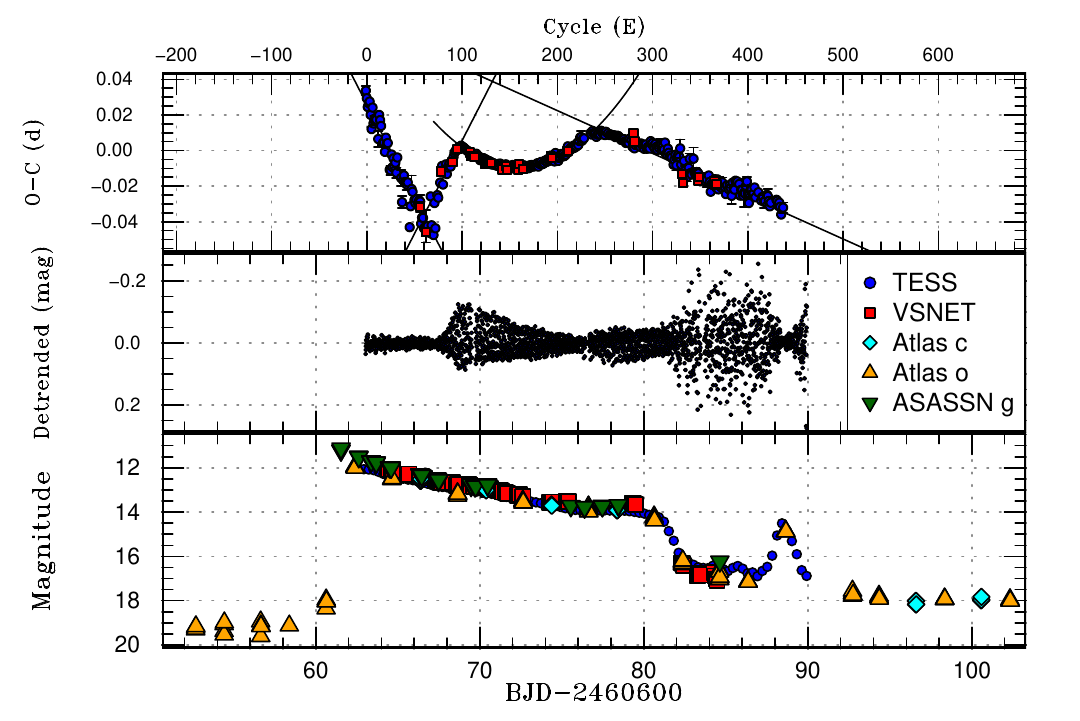}
 \end{center}
 \caption{
    Top panel; $O-C$ diagram of superhump maxima. The red squares and blue circles represent VSNET and TESS observations, respectively. $C= {\rm BJD~} 2460663.091500 + 0.058149 E$. The solid straight lines correspond to the determined periods of the early, stage-A, and stage-C superhump from left to right. The curved line shows the $P_{\rm dot}$ in stage-B superhumps.
    Middle panel; De-trended light curve of TESS observations, binned in 0.01 d.
    Bottom panel; Overall optical light curve in outburst. Red squares, blue circles, cyan diamonds, orange upper triangles, and green lower triangles represent the observations by VSNET, TESS, Atlas $c$ band, Atlas $o$ band, and ASAS-SN $g$ band, respectively. VSNET and TESS data are binned in 0.06 and 0.3 d, respectively.
    {Alt text: Three line graphs showing the evolution of light curve and superhump in outburst.}
    }
\label{fig:longterm}
\end{figure*}

\begin{figure}[tbp]
 \begin{center}
  \includegraphics[width=\linewidth]{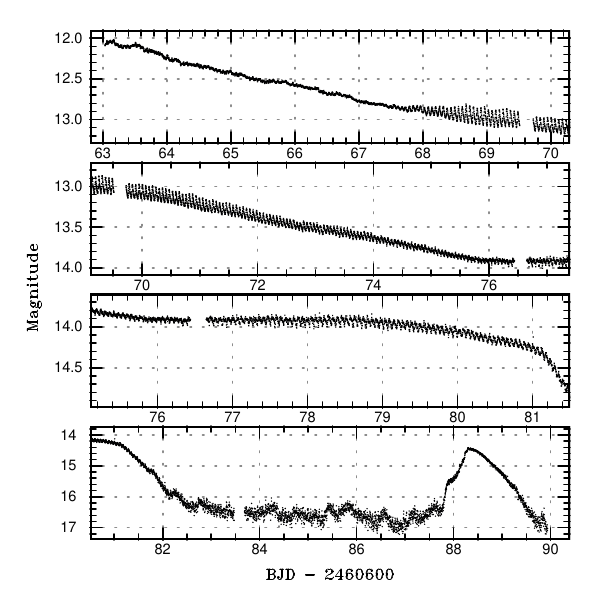}
 \end{center}
 \caption{
    Zoomed light curve of the TESS dataset during the outburst and the following rebrightening.
    {Alt text: four line graphs. The x-axises correspond to Barycentric Julian Day 2460662.8--2460670.3, 2460669.1, --2460677.4, 2460675.1--2460681.5, and 2460680.5--2460690.4, from top to bottom.}
    }
\label{fig:tesslc}
\end{figure}

The bottom panel of figure \ref{fig:longterm} presents the global light curve of ASASSN-24hd around its 2024-2025 outburst. The zoomed light curves of TESS are presented in figure 2.
The ATLAS detections on BJD 2460660.6 at $o \simeq 18.0$ mag are likely the earliest detection of ASASSN-24hd in outburst, which is $\simeq 1.0$ mag brighter than its pre-outburst level. Considering the peak magnitudes in ASAS-SN at $g = $11.120(9) mag and the Gaia quiescence counterpart (19.151(4) mag),  the outburst amplitude is $\approx$ 8.0 mag, typical of a WZ Sge-type DN \citep{kat15wzsge}. The absolute magnitude at outburst maximum $M_{g \rm - max}$ is $3.6(1)$ mag without considering Galactic extinction. Provided the outburst peaked likely around the ASAS-SN discovery on BJD 2460661.5 and the start of rapid decline on BJD 2460681.0, the outburst duration is $\simeq 19.5$ d.
Around BJD 2560676.0, the light curve stopped declining and showed a slight brightening trend. This flat phase lasted $\approx$ 5.0 d until the rapid decline. Such a halt of the outburst decline has been reported in SU UMa-type DNe with long ($\geq$ 2 years) superoutburst cycles \citep{kat03hodel, bab00v1028cyg} and in a limited number of WZ Sge-type DNe (\citet{tam23v627peg} and references therein).
The decline timescale from the outburst maximum and before this halt of decline is 7.40(1) d mag$^{-1}$ via the linear regression of the light curve, located at the lower end of the decline timescales of WZ Sge-type DNe \citep{Pdot6}.

After the rapid decline from the outburst, ASASSN-24hd underwent a single rebrightening peaking around  BJD 2460688.3, with an amplitude of $\approx$ 2.0 mag and a duration of $\approx$ 2.0 d, classifying a type-C (single and short) rebrightening episode \citep{ima06tss0222}. 
The latest measurement of ASASSN-24hd on BJD 2460717  is $o \simeq 18.5$ mag in ATLAS, still brighter than its pre-outburst level.
A shoulder at the rise of the rebrightening was observed around BJD 2460687.8. The other rebrightenings observed by Kepler or TESS, however, do not exhibit a similar shoulder at the rebrightening rise in our best knowledge (e.g., \cite{kat13j1939v585lyrv516lyr,pic21tacos,liu23tesssuuma,liu24susuma}). Between the rapid decline and the rebrightening rise, ASASSN-24hd exhibited a series of low-amplitude and short-lived "mini rebrightenings", with an interval of 0.3--0.5 d. A similar pattern is superposed during the rapid decline around BJD 2460682.2. These mini rebrightenings phenomenologically resemble those observed in V585 Lyr and K2BS5 with Kepler \citep{kat13j1939v585lyrv516lyr, boy24k2bs5} and V844 Her with TESS \citep{kat22v844her}.
\citet{mey15suumareb} interpreted these mini-rebrightenings as the consequence of a mini-version of the disk instability occurring only within the cool optically thick disk.

Lastly, we examined the available light curves of ASASSN-24hd in ATLAS, ASAS-SN, and Gaia Alert from June 2014 to this 2024--2025 superoutburst. No previous outburst is recorded, suggesting that the superoutburst cycle of ASASSN-24hd is longer than a decade.
Before the outburst rise, ATLAS had detected ASASSN-24hd at quiescence level for $> 30$ d with 2--4 d cadence. \citet{otu16DNstats} show that normal outbursts in SU UMa-type DNe peaks $\sim 0.7$-mag fainter than superoutbursts. Thus if ASASSN-24hd underwent a precursor normal outburst, it should peak $\leq 12.0$ mag. Given its decline rate $\simeq 0.7$ d mag$^{-1}$ at the rapid decline of the superoutburst and rebrightening, which is known to be similar in both normal outbursts and superoutbursts (e.g.; \cite{ham21V3101Cygrebrightening}), a precursor outburst must take $\simeq$5 days only with the decline. This conflicts with the continuous ATLAS detections. Hence we conclude that ASASSN-24hd did not show any precursor outburst.

\subsection{Superhumps}
\label{sec:3.2}

\begin{figure}[tbp]
 \begin{center}
  \includegraphics[width=\linewidth]{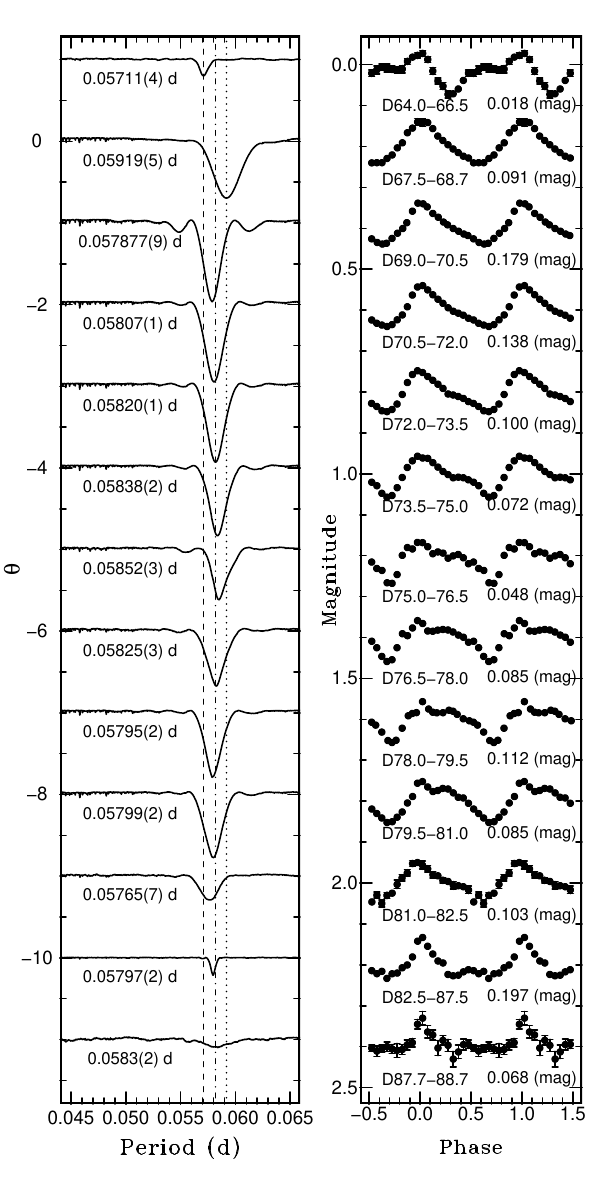}
 \end{center}
 \caption{
    Results of PDM analysis (left) and phase-folded profiles (right) in selected time windows using TESS data.
    The time window is presented below the phase-folded profile (BJD $-$ 2460600).
    The superhump periods and amplitudes are also presented below the result of each window in the left and right panels, respectively.
    The dashed, dotted, and dash-dotted lines in the left panel correspond to the periods of early, stage-A, and stage-B superhumps, respectively.
    The superhump amplitude is scaled to 0.1 mag for better visibility.
    {Alt text: two-column graphs. The best periods, time window of PDM analysis, and superhump amplitude are embedded as text in the figure. 
    }
    }
 \label{fig:pdm}
\end{figure}

TESS observations provide a precious opportunity to trace the superhump evolution in ASASSN-24hd across an outburst. 
The top panel of figure \ref{fig:longterm} shows the $O-C$ diagram of superhump maxima using $C= {\rm BJD~} 2460663.091500 + 0.058149 E$, which roughly corresponds to the best superhump period in the overall superhump stages. The times of the superhump maxima are provided in tables E3 and E4 for the VSNET and TESS datasets, respectively. The middle panel represents the normalized light curve in TESS.
Based on the $O - C$ diagram, we determined the stage-A, stage-B, and stage-C ordinary superhump phases as BJD 2460667.5--2460668.7, 2460668.9--2460676.8, and 2460677.1--2460681.0, respectively.  This stage B-C superhump transition is likely after the halt of decline around BJD 2560676.0. Before the stage-A superhumps, the $O-C$ diagram exhibits the stable and shorter period over BJD 2460663.0--2460667.0, which we interpret as the early superhump phase, characteristics of a WZ Sge-type DN (see section \ref{sec:5}).

We first determined the superhump periods solely using TESS. Through PDM analysis, we obtained \earlysh, \stageAsh, \stageBsh, and \stageCsh d for early, stage-A, stage-B, and stage-C superhumps, respectively. $P_{\rm dot}$ is determined as \pdot cycle$^{-1}$ by fitting the $O-C$ with the parabolic function during stage-B ordinary superhumps ($E=$120--220).
We note that there was no significant period (i.e. beat, WD spin) other than the above-mentioned superhumps in the TESS observations up to the available TESS sampling limit of an order of minutes (see figure \ref{fig:tessphot} as well).

Figure \ref{fig:pdm} shows the PDM and phase-folded profile of the superhumps in the selected time windows using the TESS observations. The top periodogram and profile correspond to the early superhump phase. The phase-averaged profile has double peaks with an amplitude of $\simeq 0.018$ mag, consistent with its WZ Sge-type classification in the vsnet-alert 28039. The second one, on BJD 2460667.7-2460668.7, corresponds to the stage-A ordinary superhump phase. The phase-averaged profile shows a clear, saw-shaped, single-peaked profile. The following ones correspond to stage-B ordinary superhumps up to the time window BJD 2460675.0--2460676.5. In the left panel of figure \ref{fig:pdm}, one can see the drift of the superhump period towards longer ones.  The phase-averaged profile becomes double-peaked and more flat-topped after the time window on BJD 2460673.5--2460675.0. The regrowth of superhump amplitude was observed around BJD 2460677 (see also section \ref{sec:5.3}), slightly after the halt of the declining trend. 
Between the rapid decline and the rebrightening, the superhumps showed a single-peaked profile. The superhump became undetectable on the decline of the rebrightening outburst (i.e. after BJD 2460688.5), which might be due to its low amplitude and signal-to-noise ratio rather than the physical phenomena.

Analyzing the VSNET observations independently, we obtain the stage-B superhump period as \stageBshV d and $P_{\rm dot}$ as \pdotV cycle$^{-1}$. As most of our ground-based observations during the stage-B superhumps were obtained in its first half, the inconsistency of the stage-B superhump periods between the VSNET and TESS datasets is likely due to the nonuniform VSNET data. The $P_{\rm dot}$ in stage-B superhumps is consistent between the VSNET and TESS datasets, as it determines the secure change of the superhump period over stage-B. Lastly, the superhump period on BJD 2460682--2460684, corresponding to the mini-rebrightening phase, is determined as \lateshV d using the VSNET dataset.

\section{Optical spectra}
\label{sec:4}

\begin{figure}[tbp]
 \begin{center}
  \includegraphics[width=\linewidth]{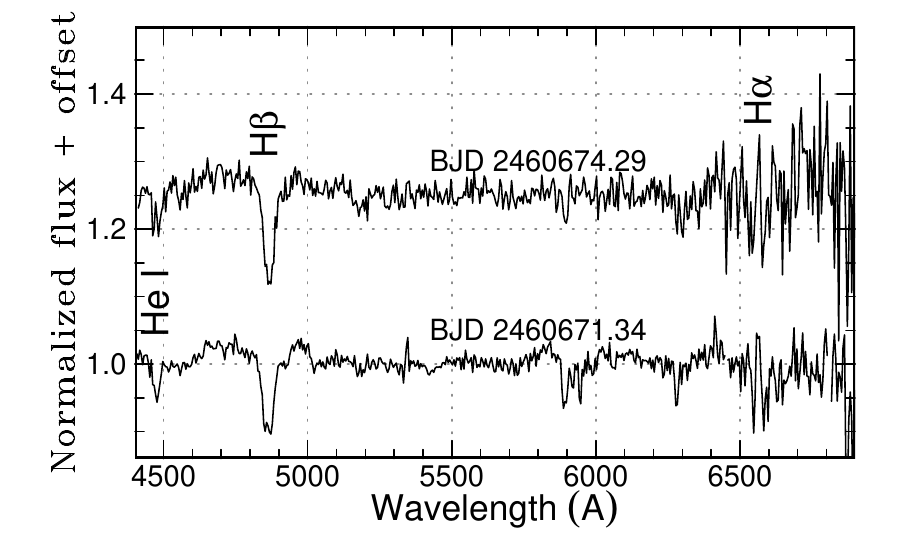}
 \end{center}
 \caption{
    Normalized spectra of ASASSN-24hd on BJD 2460671.34 (lower) and 2460674.29 (upper), observed with the 1.9m telescope and the SpUpNIC at the Sutherland observatory.
    {Alt text: one line graph.}
    }
    \label{fig:spec}
\end{figure}

Figure \ref{fig:spec} presents the normalized spectra of ASASSN-24hd in outburst on BJD 2460671.34 (lower) and 2460674.29 (upper). These epochs correspond to the stage-B superhump phase.
Both optical spectra are characterized by the absorption lines of He \textsc{i}~4471, H$\beta$, and H$\alpha$. H$\alpha$ accompanies the narrow emission component. By fitting with Gaussian function to H$\beta$, the full-width half max and equivalent width are obtained as 46(2) and -6.0(9) \AA~ on BJD 2460671.34, and 42(2) and -7.0(8) \AA~ on BJD 2460674.29, respectively.

\section{Discussion}
\label{sec:5}

As presented in section \ref{sec:3}, our detection of early superhumps in ASASSN-24hd establishes its classification as a WZ Sge-type DN. However, we find that its outburst light curve and superhump evolution share a great similarity to those of V585 Lyr observed with Kepler, of which \citet{kat13j1939v585lyrv516lyr} discussed that this system is an SU UMa-type DN. Thus we first present other supporting facts on our classification in section \ref{sec:5.1}. Then, we discuss the possible outburst scenarios explaining their similarities and differences (section \ref{sec:5.2}). We finally give a detailed view of its superhump evolution in section \ref{sec:5.3}.

\subsection{ASASSN-24hd as a WZ Sge-type dwarf nova}
\label{sec:5.1}

\begin{figure}[tbp]
 \begin{center}
  \includegraphics[width=\linewidth]{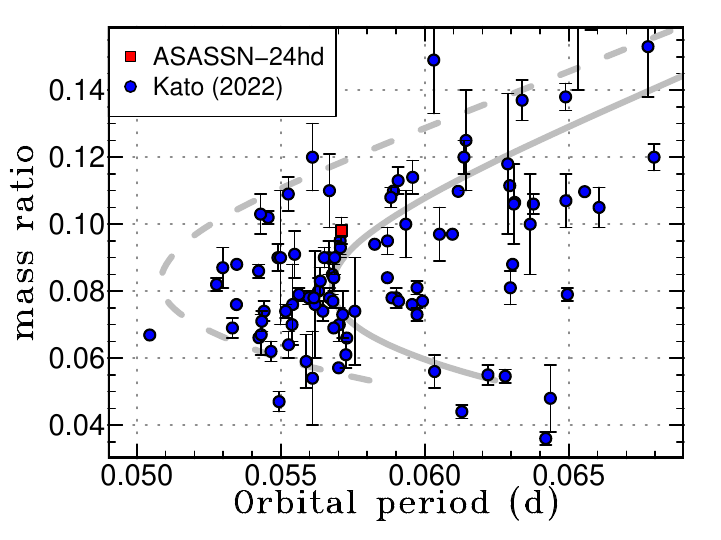}
 \end{center}
 \caption{
    The orbital period vs. the mass ratio relation in short-$P_{\rm orb}$ CVs. The red square and blue circles represent those of ASASSN-24hd and other CVs from \citet{kat22updatedSHAmethod}, respectively. The dashed and solid lines show the theoretical and empirical evolutionary tracks, respectively, from \citet{kni11CVdonor}.
    {Alt text: one line graph depicting the evolutionary state of ASASSN-24hd and WZ Sge-type DNe.}
    }
 \label{fig:evolution}
\end{figure}

In this subsection, we summarize our classification of ASASSN-24hd as a WZ Sge-type DN. As introduced in section \ref{sec:1}, the most accepted classification criterion of a WZ Sge-type DN is the detection of early superhumps before the appearance of ordinary superhumps. Our time-resolved observations by TESS detect a double-peaked profile with a stable period of \earlysh d, shorter than those of the following ordinary superhumps. Its duration is at least $> 4.0$ d and $> 50$ cycles solely from TESS, and is $\simeq 5.5$ d considering the light curve maximum observed by ASAS-SN since early superhumps appear around the outburst maximum \citep{ish02wzsgeletter}.
This is much longer than that of the pre-stage-A oscillations reported in \citet{kat22v844her, boy24k2bs5}. 
The observed maxima during this phase correspond to the epoch expected from the maximum in the phase-averaged profile. Thus, we interpret this modulation as early superhumps and classify ASASSN-24hd as a WZ Sge-type DN. 
The early superhump amplitude of $\approx 0.018$ mag suggests that ASASSN-24hd is a middle inclined system, approximately 50--55$^\circ$ according to \citet{kat22WZSgecandle}. This is consistent with our optical spectra in outburst, dominated by Balmer absorption lines and not exhibiting any strong Balmer nor He~\textsc{ii} emission lines \citep{tam21seimeiCVspec}.

With this interpretation and regarding the early superhump period as the orbital one, we applied the relation between the superhump excesses in the stage-A superhumps and mass ratios in \citet{kat13qfromstageA, kat22updatedSHAmethod}. This yields the mass ratio of ASASSN-24hd as \massratio. Furthermore, using the $P_{\rm dot}$ during the stage B superhumps and equation 6 in \citet{kat15wzsge}, the mass ratio of ASASSN-24hd is estimated as 0.11(2), consistent with the above value.
Figure \ref{fig:evolution} presents the relation of the orbital (or early superhumps) periods $P_{\rm orb}$ and the mass ratios $q$ in ASASSN-24hd (red square) and short $P_{\rm orb}$ CVs (blue circles; \cite{kat22updatedSHAmethod}). ASASSN-24hd is located above the period minimum, following the empirical evolution track in \citet{kni11CVdonor}. The mass ratio of ASASSN-24hd satisfies the upper limit of $q \lesssim 0.1$ observed in WZ Sge-type DNe \citep{kat22updatedSHAmethod}.  Its classification as a potential WD in \citet{gaiaDR3syntphot} supports that the WD dominates its quiescence brightness.
Lastly, its outburst amplitude (8.0 mag), duration (19.5 d), lack of a precursor outburst, and superoutburst cycle exceeding 10 years are all comparable to other WZ Sge-type DNe \citep{kat15wzsge}. These points strongly support our classification of ASASSN-24hd as a rather normal WZ Sge-type DN with a relatively large mass ratio.

\subsection{The 2024 ASASSN-24hd vs 2010 V585 Lyr superoutbursts}
\label{sec:5.2}

\begin{figure}[tbp]
 \begin{center}
  \includegraphics[width=\linewidth]{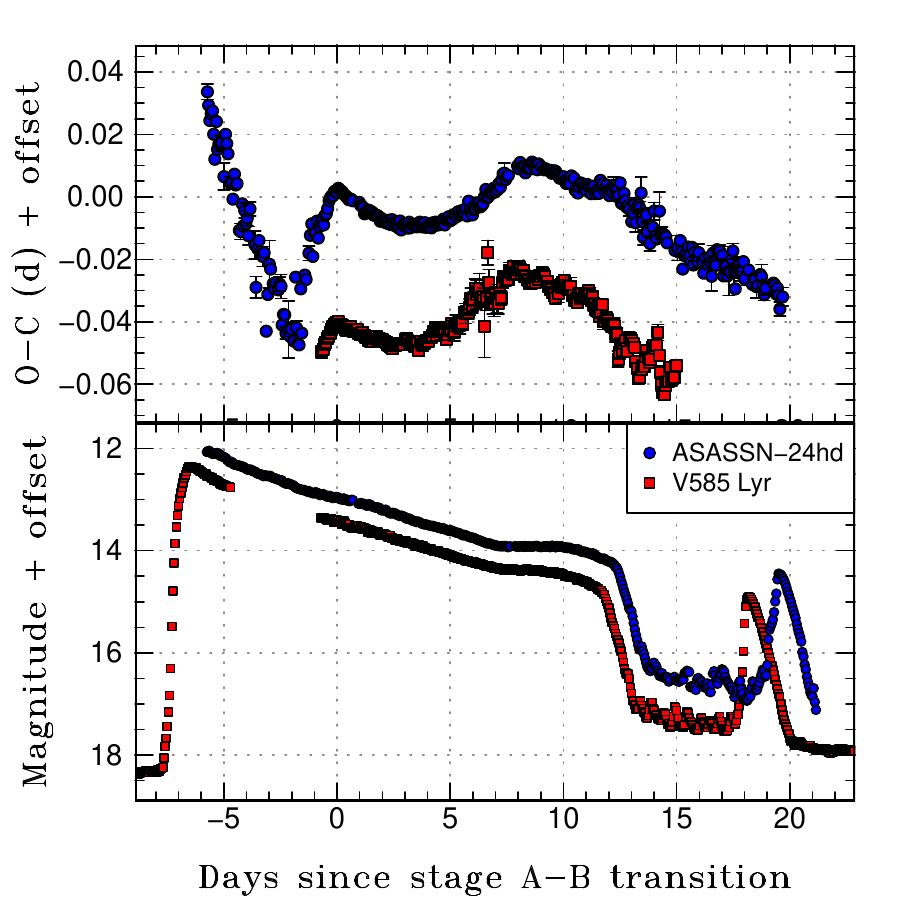}
 \end{center}
 \caption{
    The comparison of the $O-C$ diagrams (top) and light curves (bottom) of ASASSN-24hd (blue circles) and V585 Lyr (red squares; \cite{kat13j1939v585lyrv516lyr}) using the TESS and Kepler data.
    The zero point of the x-axis is normalized at the epoch corresponding to the stage-A to stage-B superhump transition.
    {Alt text: two line graph comparing V585 Lyr and ASASSN-24hd in outburst. 
    }
    }
    \label{fig:occomp}
\end{figure}

\begin{table}
\caption{Comparison of ASASSN-24hd and V585 Lyr.}
\centering
\label{tab:1}
\begin{tabular}{ccc}
  \hline     
       & ASASSN-24hd & V585 Lyr  \\
    year   & 2024 & 2010  \\
    \hline
    stage-B $P_{\rm SH}$ (d) & \stageBsh & 0.06041(1)\commenta \\
    stage-B $P_{\rm dot}$ (cycle$^{-1}$)& \pdot & 9.6(5) $\times$ 10$^{-5}$\commenta \\
    $q$ from Stage-A $P_{\rm SH}$ & \massratio & -- \\
    $q$ from Stage-B $P_{\rm dot}$ & 0.11(2) & 0.10(2) \\
    Distance (pc)\commentb & 318.9$^{+18.6}_{-17.8}$ & -- \\
    $M_{\rm max}$ (mag) & 3.6(1) &  -- \\
    $M_{G- \rm min}$ (mag)\commentc & 11.6(1) & -- \\
    Amplitude (mag) & 8.0 & 6.0 \\
    Duration\commentd (d) & 19.5 & 18.4\commenta \\
    Outburst cycle (yr) & $>$ 10 & $\simeq$ 2.3\commente \\
  \hline
    \multicolumn{3}{l}{\commenta \citet{kat13j1939v585lyrv516lyr}.}\\
    \multicolumn{3}{l}{\commentb \citet{Bai21GaiaEDR3distance}.}\\
    \multicolumn{3}{l}{\commentc \citet{gaiaedr3}.}\\
    \multicolumn{3}{l}{\commentd From outburst maximum to beginning of rapid decline.}\\
    \multicolumn{3}{l}{\commente \citet{Pdot8}.}\\
\end{tabular}
\end{table}

We have found that the outburst light curve profile and superhump evolution of ASASSN-24hd share a great similarity with those of the 2010 superoutburst of V585 Lyr studied with Kepler \citep{kat13j1939v585lyrv516lyr}. Figure \ref{fig:occomp} compares their light curves and $O-C$ diagrams in outburst, where the x-axis is normalized at the epoch when the stage A-B superhump transition occurs. Table \ref{tab:1} summarizes the key numbers of ASASSN-24hd and V585 Lyr in outburst and quiescence.
We note that there is no available parallax and distance of V585 Lyr in Gaia EDR3.
Both superoutbursts lack a precursor outburst, and show similar outburst duration, decline timescales, long waiting time before the stage A-B superhump transition, halt of declining $\approx 5$ d before the rapid decline, mini-rebrightenings, and single-short rebrightening. 
The $P_{\rm sh}$ and $P_{\rm dot}$ of the stage-B superhumps in V585 Lyr are 0.06041(1) d and 9.6(5) $\times$ 10$^{-5}$, respectively. Its mass ratio of 0.10(2) based on this $P_{\rm dot}$ (the equation 6 in \citet{kat15wzsge}) suggests that the evolutionary state of ASASSN-24hd and V585 Lyr can be similar, although the mass ratio based on stage-B superhump $P_{\rm dot}$ tends to underestimate \citep{kat22updatedSHAmethod}.

However, although there is an observation gap in the early stage of the 2010 superoutburst of V585 Lyr in Kepler data, time-resolved observations in its other superoutbursts have never reported early superhumps (\cite{Pdot8} and references therein). \citet{kat13j1939v585lyrv516lyr} interpreted its lack of a precursor outburst and long waiting time before the stage A--B superhump transition as the type B superoutburst in \citet{osa03DNoutburst, osa05DImodel}; the mass stored at the onset of the superoutburst is large enough for the disk to reach the tidal truncation radius beyond the 3:1 resonance radius (but below the 2:1 resonance radius) and to sustain the disk in outburst until the tidal instability develops.

Inspired by its similarity to ASASSN-24hd in outburst, another possible explanation for the lack of early superhumps in V585 Lyr is a face-on disk. Early superhumps are proposed to originate from the geometrical effect of a vertically-extended spiral-arm structure in a disk \citep{osa02wzsgehump, uem12ESHrecon}. Hence one expects that early superhump amplitudes become below the detectable limit for face-on systems, below an inclination of 40$^{\circ}$ according to \citet{kat22WZSgecandle}.
Meanwhile, the superoutburst cycle of V585 Lyr is $\simeq$ 2.3 years, much shorter than those typically found in WZ Sge-type DNe and ASASSN-24hd \citep{kat15wzsge}. V585 Lyr showed a precursor outburst in its 1967 superoutburst \citep{kry01v585lyrv587lyr} and even a normal outburst without any following superoutburst in 2013 in Kepler \citep{Pdot8}, which are characteristics of SU UMa-type DNe with higher mass-transfer rate.
Moreover, the outburst amplitude of ASASSN-24hd is 2.0-mag larger than that of V585 Lyr despite the face-on disk required to explain the lack of early superhumps in V585 Lyr. 
Hence, V585 Lyr cannot be interpreted simply as a twin of ASASSN-24hd with a lower inclination.

Given the similar orbital period and mass ratio between ASASSN-24hd and V585 Lyr, the only feasible binary parameters that may differ are the mass transfer rate and properties of the primary WD.
Indeed, a long superoutburst cycle and a lack of normal outbursts in WZ Sge-type DNe are invoked to (1) very low disk viscosity in quiescence, an order of magnitude lower than that of SU UMa-type DNe, probably due to lower mass-transfer rate \citep{sma93wzsge, osa95wzsge} or (2) removal of the inner disk by the WD magnetosphere or the evaporation effect \citep{ham97wzsgemodel, mey98wzsge, mat07wzsgepropeller}. 
Both studies have also shown that these factors affect the angular momentum distribution in a disk. \citet{osa02wzsgehump} gave the upper range of the mass ratio of WZ Sge-type DNe by comparing the 2:1 resonance and the maximum disk radius in outburst in the case where the transferred mass accumulates at the circularization radius due to its extremely low viscosity. \citet{mat07wzsgepropeller} showed that, while the quiescence viscosity of WZ Sge remains similar to that of SU UMa-type DNe, the inner disk truncation significantly extends the superoutburst cycles and triggers an outburst at larger disk radii with higher critical surface density, resulting in all outbursts in superoutburst. The suppression of the growth of the 3:1 resonance by the 2:1 resonance \citep{lub91SHa, lub91SHb} may prolong the early superhump phase in ASASSN-24hd even with a less massive disk. Therefore, their differences in outburst behaviors seem to be strongly linked to their quiescence properties, although further studies in quiescence are required to conclude.

An important aspect to distinguish these scenarios is the magnetic WD population in CVs. A larger magnetosphere is expected in a system with a lower accretion rate (i.e. WZ Sge-type DNe in quiescence), enabling inner disk truncation. There have been a handful number of WZ Sge-type DNe where its magnetic nature is confirmed; e.g. WZ Sge and V455 And \citep{pat81wzsge, ara05v455and}. More recently, \citet{pav19a19fk, kol25gaia19cwm, cas25goto0650} detected a probable spin period after the rapid decline of the superoutburst in WZ Sge-type DNe, suggesting that they host a magnetized WD.  \citet{ver24awdqpo} detected the quasi-period oscillations (QPOs) using TESS data of CVs in quiescence including some WZ Sge-type DNe. They interpreted that these QPOs originate from the magnetically driven warping of a disk.  Although we did not find any suspicious spin period in the available data of ASASSN-24hd in outburst and post-outburst decline, since ASASSN-24hd will be covered by the TESS in the following sectors up to Sector 90 before returning to quiescence, additional datasets and analysis of these TESS data may provide evidence of a magnetic WD in ASASSN-24hd.

Future statistical studies to check the fraction of superoutbursts with and without early superhumps at a wide range of mass ratios and orbital (or superhump) periods may also benefit from distinguishing these scenarios. If V585 Lyr and similar systems are indeed a face-on WZ Sge-type DN, the fraction of superoutbursts without detectable early superhump (i.e.; fraction of face-on systems) at mass ratio $q \simeq 0.1$ should be the same as more evolved WZ Sge-type DNe; for example, period bouncers. In the other case, the fraction of superoutbursts without early superhump should be higher (face-on systems plus superoutbursts intrinsically lacking an early superhump phase) at mass ratios $q \simeq 0.1$ compared to the more evolved WZ Sge-type DNe.

We finally note here that the above discussion may not be valid if the outbursts of ASASSN-24hd and V585 Lyr are both maintained by the enhanced mass transfer from the irradiated secondary \citep{ham97wzsgemodel, ham20CVreview}. In this case, since most of the accreting gas during an outburst is supplied by the enhanced mass transfer, the disk state before the mass transfer burst will affect only the length of superoutburst cycles, but not very much the outburst profile or superhump evolution. Thus, their similar outburst duration requiring a similar disk mass conflicts with the absence of early superhumps only in V585 Lyr.

\subsection{Superhump evolution}
\label{sec:5.3}

\begin{figure*}[tbp]
 \begin{center}
  \includegraphics[width=\linewidth]{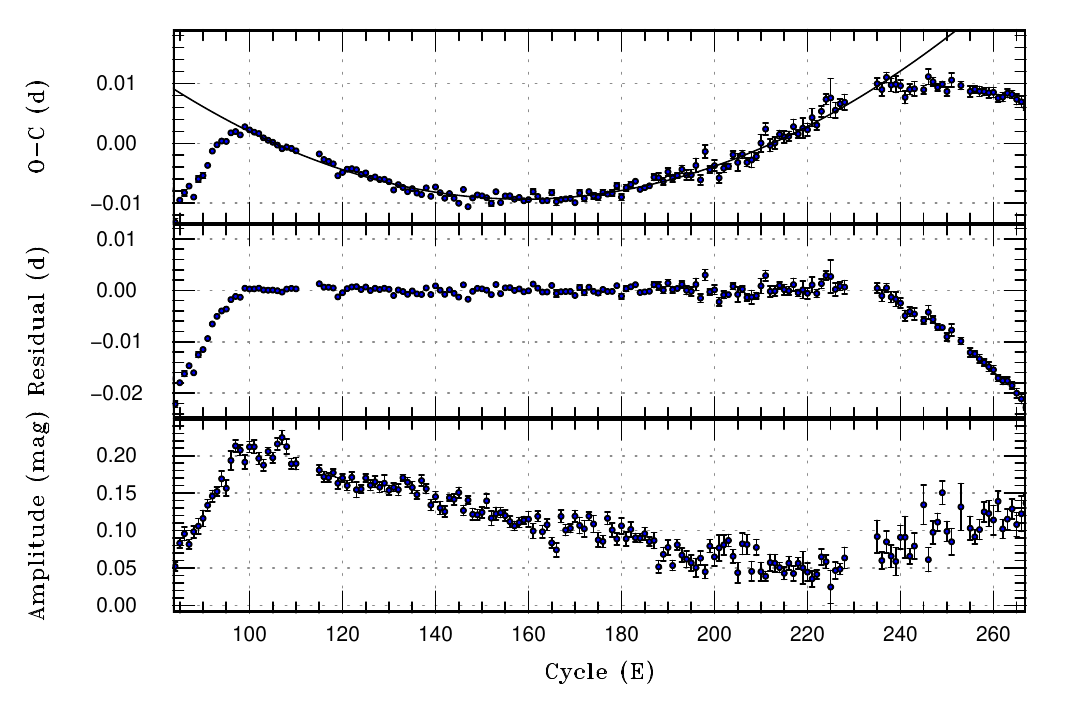}
 \end{center}
 \caption{
    Parabolic fit to the $O-C$ during the stage-B ordinary superhumps (top) and its residual (bottom). The bottom panel represents the evolution of superhump amplitude on a magnitude scale.
    {Alt text: Three-line graph depicting the zoomed $O-C$ diagram and amplitude evolution. }
    }
    \label{fig:ocstgAB}
\end{figure*}

In this subsection, we discuss the $O-C$ evolution of ordinary superhumps in detail. TESS provides uninterrupted observations over the entire outburst of ASASSN-24hd, which is ideal for examining the superhump evolution in a WZ Sge-type DN. The only previous such observation is KSN:BS-C11a observed by Kepler \citep{rid19j165350}, although its 30-min cadence prevents detection of individual superhump maxima. Unfortunately, the early superhump amplitude was too small to examine the transition from the early to ordinary superhumps. WZ Sge and V455 And are the better examples of this transition \citep{Pdot, kat15wzsge}. The top panel of figure \ref{fig:ocstgAB} presents the zoomed $O-C$ diagram around the stage-B superhumps. The middle panel shows its residual from the parabolic fit (the solid line in the upper panel). The bottom panel presents the superhump amplitude on a magnitude scale. In general, we confirmed that the stage transition in ordinary superhumps in ASASSN-24hd is consistent with that of the SU UMa-type DNe observed by Kepler and TESS.

Figure 22 in \citet{kat15wzsge} gives a relation in WZ Sge-type DNe between the duration of stage-A superhumps and $P_{\rm dot}$ in stage-B superhumps. Those of ASASSN-24hd ($\simeq$ 20 cycles and \pdot cycle$^{-1}$) are one of the shortest durations and largest $P_{\rm dot}$, aligning with their relation. At the stage A--B superhump transition, the superhump maxima showed a smooth transition; i.e. there was no jump of the phase of the superhump maxima. As seen in the $O-C$ diagram, the constant period of the stage-A superhumps continues up to $E=97$, and the stage-B superhump probably starts from the cycle $E=99$. We cannot judge which stage the cycle $E=98$ belongs to. The superhump amplitude reached its maximum over $E=$97--108, about the same time as the superhump stage transition.

Thus, the timescale of the stage A-B superhump transition is $\simeq$ 2--3 orbital cycles. Such a fast and abrupt transition seems to be common in short-$P_{\rm sh}$ superoutbursts if the coverage is high enough (e.g. V585 Lyr in \cite{kat13j1939v585lyrv516lyr}; V844 Her in \cite{kat22v844her}), while the transition is more smooth in some systems with longer orbital periods (e.g. V344 Lyr in \cite{Pdot3}; V516 Lyr in \cite{kat13j1939v585lyrv516lyr}). 
In the TTI model, the stage A--B superhump transition is understood as the onset of propagation of eccentricity from the 3:1 resonance radius to the inner disk. Although the pressure effect is regarded as negligible in the stage-A superhump phase, this should play a more prominent role once the stage-B superhump phase begins, especially in low-mass ratio systems \citep{osa13v344lyrv1504cyg}.  \citet{kat13j1939v585lyrv516lyr} discussed that since the pressure effect decreases the precession rate, a more abrupt period decrease in systems with a lower mass ratio can be attributed to the relative importance of the pressure effect on the precession rate. Recent simulation works by \citet{jor24ttisimulation} also showed the rapid (in a few $P_{\rm orb}$) decrease of the disk precession rate after the development of superhumps.

There are only a handful of WZ Sge-type DNe showing stage-C superhumps (e.g. V1251 Cyg, V748 Hya; \cite{Pdot}), while stage-C superhumps are more commonly observed in SU UMa-type DNe \citep{Pdot}.  Based on the $O-C$ diagram, ASASSN-24hd is another unambiguous example of a WZ Sge-type DN showing stage-C superhumps, likely lasting up to the beginning of the rebrightening. Compared to the stage A--B superhump transition, the stage B--C superhump transition is less clear in our data particularly because of the observation gap around BJD 24606776.5 in TESS. 
Nevertheless, we did not find any phase jump of the primary superhump maxima from those of the single-peaked superhumps to the double-peaked superhump profiles (see BJD 2460673.5--2460681.0 in figure \ref{fig:pdm} as well). The stage-B superhumps continue up to at least $E=237$, then the stage-C superhumps likely started $E\simeq 240$. The regrowth of superhump amplitude probably occurs at the same time, but after the halt of decline ($E\simeq 220$), as observed in other systems with a halt of declining trend before the rapid decline \citep{kat03hodel}. Hence the stage B--C superhump transition timescale seems to be also $\lesssim 5$ cycles, while the larger scatter in the $O-C$ diagram prevents us from confirming that. As the superoutburst enters the rapid decline, the second maxima diminish and the superhump profile returns to single-peaked. These respects agree with those observed in V585 Lyr and other superoutbursts in Kepler and TESS.

The origins of the stage B--C superhump transition and the growth of a double-peaked profile are still unclear. However, it is worth noting that the shorter and fainter superoutbursts lacking early superhump phase in WZ Sge-type DNe EG Cnc, AL Com, and V3101 Cyg did not show any clear stage-C superhumps \citep{kim16alcom, kim21EGCnc, tam20j2104}, while V627 Peg showed the stage B--C superhump transition in all of its superoutbursts, regardless of the presence of the early superhump phase \citep{tam23v627peg}. Thus the occurrence of clear stage B--C superhump transition seems sensitive to the mass ratio, rather than the occurrence of the 2:1 resonance.

\section{Summary}
\label{sec:6}

We present time-resolved observations of a new WZ Sge-type DN ASASSN-24hd in its 2024-2025 superoutburst, together with the coincident TESS observations. The key findings are summarized as follows;

\begin{itemize}
    \item 
        Its superoutburst amplitude and duration were 8.0 mag and 19.5 d, respectively. Based on the non-detection of any previous outbursts in available time-domain surveys, the superoutburst cycle likely exceeds a decade. The superoutburst showed a halt of decline $\approx$ 5 d before the rapid decline, typical of superoutbursts in DNe with a mass ratio of $q \simeq 0.1$. Our TESS observations detect early superhumps, confirming its classification as a WZ Sge-type DN. The early and stage-A ordinary superhump periods are determined as \earlysh and \stageAsh d, respectively. The $P_{\rm dot}$ in stage-B superhumps is determined to be \pdot cycle$^{-1}$ using the TESS data. These superhump periods yield the mass ratio of ASASSN-24hd as \massratio. We also find that ASASSN-24hd is a rare WZ Sge-type DN showing the stage-C superhumps, with a period of \stageCsh d.

    \item 
        The continuous observations in TESS have provided the most detailed view of the evolution of superhumps in WZ Sge-type DNe. The stage A-B and B-C superhump transition occurs within a few orbital cycles. There is no phase jump of maxima during the ordinary superhump phase, while double-peaked profiles are observed between the latter half of the stage-B superhump phase and the rapid decline of the outburst. Before the stage B--C superhump transition and the regrowth of superhump amplitude, the halt of outburst decline is observed. 
        We confirm that these observed features in ASASSN-24hd are generally consistent with those of short-$P_{\rm sh}$ SU UMa-type DNe covered by Kepler and TESS. 

    \item 
        The outburst light curve and $O-C$ diagram of superhump periods of ASASSN-24hd well resemble those of the 2010 superoutburst in V585 Lyr observed by Kepler including the long waiting time before the stage A--B transition, which \citet{kat13j1939v585lyrv516lyr} interpreted as a result of large mass accumulated in quiescence rather than the excitement of the 2:1 resonance. 
        The straightforward interpretation of V585 Lyr being a low-inclination WZ Sge-type, however, disagrees with the characteristics of V585 Lyr compared to ASASSN-24hd; its $\simeq$ 2.3-year superoutburst cycle and smaller outburst amplitude. Instead, we propose that either an extremely low quiescence viscosity or a truncation of the inner disk, invoked to explain the extreme nature of WZ Sge-type DNe, may enable the disk to reach the 2:1 resonance radius in ASASSN-24hd.

\end{itemize}

\begin{ack}

We acknowledge amateur and professional astronomers around the world who have shared data on variable stars and transients with the VSNET collaboration. This paper uses observations made from the South African Astronomical Observatory (SAAO)

The IRSF project is a collaboration between Nagoya University and the South African Astronomical Observatory (SAAO) supported by the Grants-in-Aid for Scientific Research on Priority Areas (A) (No. 10147207 and No. 10147214) and Optical \& Near-Infrared Astronomy Inter-University Cooperation Program, from the Ministry of Education, Culture, Sports, Science and Technology (MEXT) of Japan and the National Research Foundation (NRF) of South Africa. We thank Dr. Takayoshi Kusune for supporting our observations with IRSF. 

This research made use of Lightkurve, a Python package for Kepler and TESS data analysis \citep{lightkurve}, which is implemented based on Astropy \citep{astropyv5}, Astroquery \citep{astroquery}, and tesscut \citep{bra19tesscut}.
This work has made use of data from the Asteroid Terrestrial-impact Last Alert System (ATLAS) project. The ATLAS project is primarily funded to search for near earth asteroids through NASA grants NN12AR55G, 80NSSC18K0284, and 80NSSC18K1575; byproducts of the NEO search include images and catalogs from the survey area. This work was partially funded by Kepler/K2 grant J1944/80NSSC19K0112 and HST GO-15889, and STFC grants ST/T000198/1 and ST/S006109/1. The ATLAS science products have been made possible through the contributions of the University of Hawaii Institute for Astronomy, the Queen’s University Belfast, the Space Telescope Science Institute, the South African Astronomical Observatory, and The Millennium Institute of Astrophysics (MAS), Chile.
We acknowledge ESA Gaia, DPAC and the Photometric Science Alerts Team (http://gsaweb.ast.cam.ac.uk/alerts).

\end{ack}

\section*{Supporting Information}
The following Supporting Information is available in the online version of this article: Tables E1--E4.


\bibliographystyle{pasjtest}
\bibliography{cvs}


\appendix

\renewcommand{\thetable}{A\arabic{table}}
\renewcommand{\thefigure}{A\arabic{figure}}

\section*{TESS photometry of ASASSN-24hd}
\label{sec:tessphot}

\begin{figure}[tbp]
 \begin{center}
  \includegraphics[width=\linewidth]{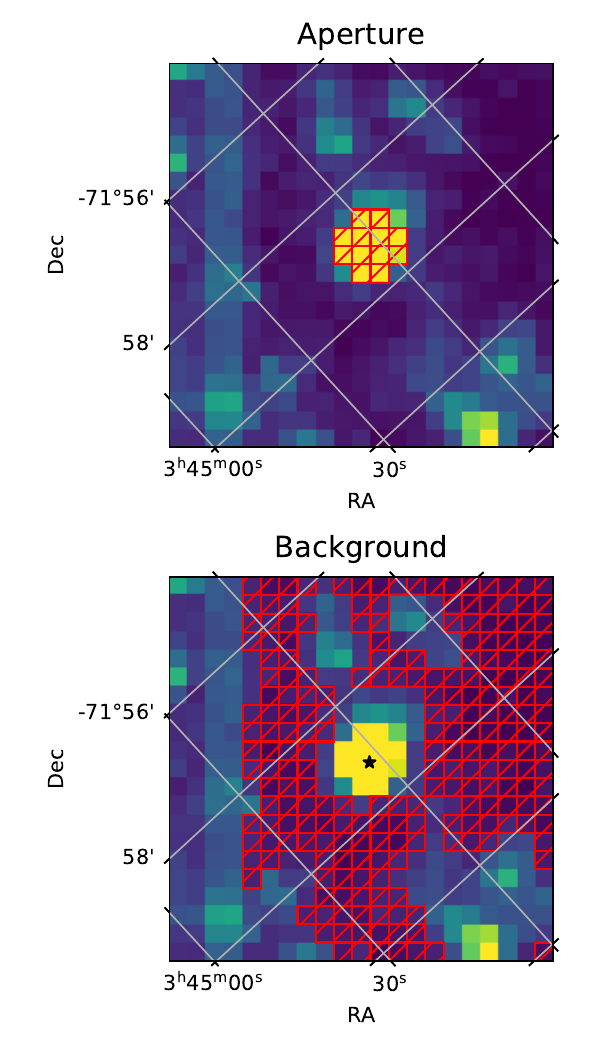}
 \end{center}
 \caption{
    The red-hatched pixels show those used for the target aperture (top) and background estimation (bottom) in our TESS aperture photometry. 
    The black star in the bottom panel shows the position of ASASSN-24hd.
    {Alt text: two-line graph illustrating the 21 $\times$ 21 pixels around ASASSN-24hd in TESS.}
    }
    \label{fig:tessaper}
\end{figure}

\begin{figure}[tbp]
 \begin{center}
  \includegraphics[width=\linewidth]{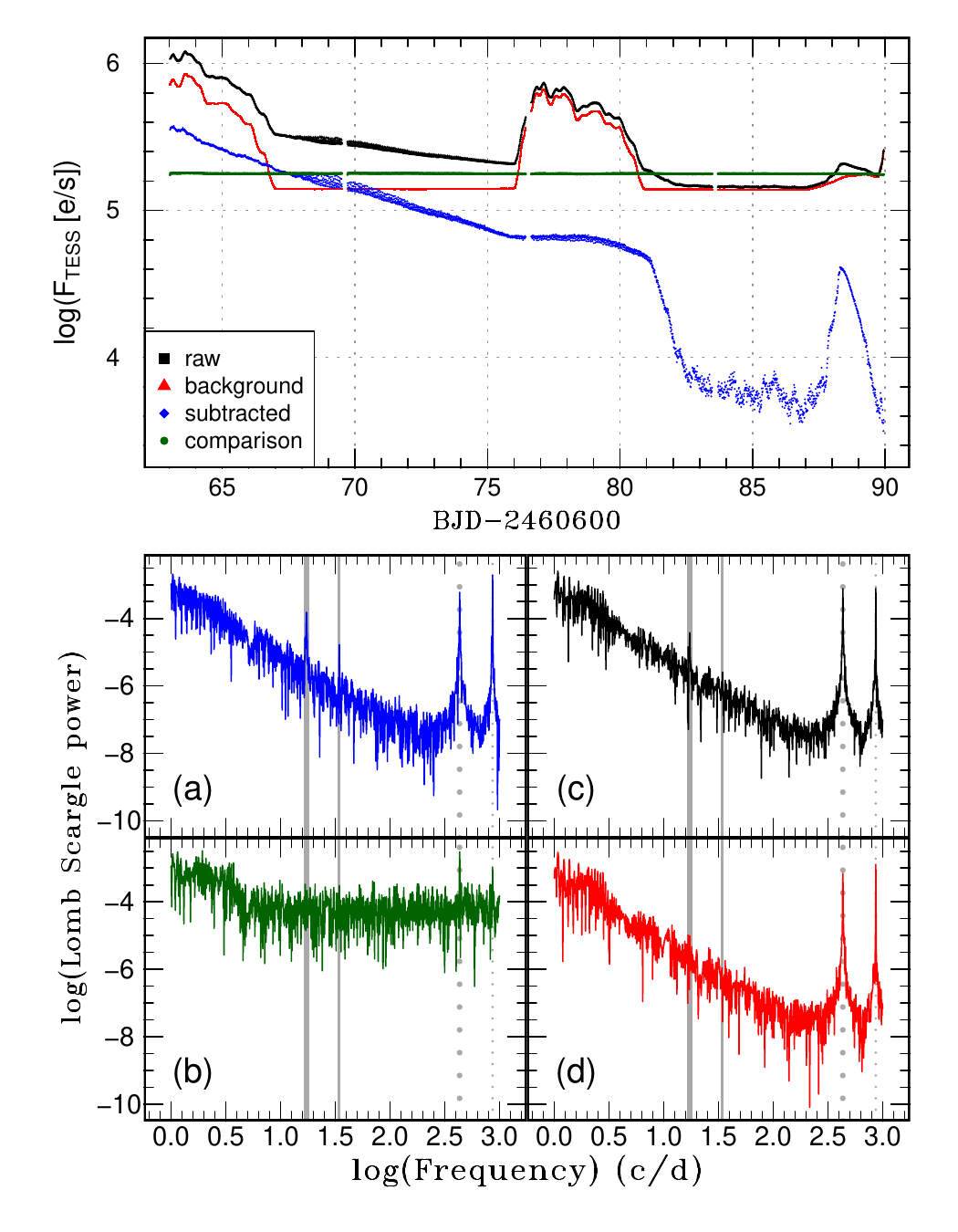}
 \end{center}
 \caption{
    Top panel; the raw (black), background (red), and subtracted (blue) light curve in TESS flux $F_{\rm TESS}$ unit. The green light curve represents the subtracted one of the nearby standard star TIC 238193289. The light curves are binned in 0.01 d.
    Bottom panels; the Lomb-Scargle results in the frequency range 1--1000 d$^{-1}$ of the raw (c), background (d), and subtracted (a) light curves. The same analysis was performed on the standard star TIC 238193289 (b). The solid and dotted vertical lines represent the frequencies corresponding to the stage-B superhumps and TESS observation cadence, respectively. The thinner lines show the double frequency of them.
    {Alt text: five panels showing that our aperture photometry of TESS does not suffer from any background contamination or false signals.}
    }
    \label{fig:tessphot}
\end{figure}

As only the TICA FFIs are available at the time of submission, we performed the photometry of ASASSN-24hd using the \texttt{lighukurve} package \citep{lightkurve} by ourselves.
Figure \ref{fig:tessaper} shows the pixels we used for the aperture photometry of ASASSN-24hd (upper) obtained via  \texttt{create$\_$threshold$\_$mask} in \texttt{lightkurve.TessTargetPixelFile} using the first 200 images of the sector 87 and threshold of 7.0, and the background estimation using \texttt{estimate$\_$background} in \texttt{lightkurve.TessTargetPixelFile} (lower).
Figure \ref{fig:tessphot} shows the raw aperture flux extracted by \texttt{extract$\_$aperture$\_$photometry} in \texttt{lightkurve.TessTargetPixelFile} (black), estimated background flux (red), and subtracted flux (blue) in the TESS flux scale (e s$^{-1}$).
The green light curve shows that of the AAVSO standard star TIC 238193289 (=AUID 000-BQB-801, $V=12.95(6)$ mag), calibrated in the same manner as ASASSN-24hd.
The panels (a)--(d) in figure \ref{fig:tessphot} present the result of the Lomb-Scargle analysis of them. The dashed line represents the frequency of stage-B superhumps, which is only significant in the subtracted and raw light curves of ASASSN-24hd.
This confirms that our subtracted light curve and periodogram of ASASSN-24hd are not contaminated with any background trends.
We converted the subtracted TESS flux $F_{\rm TESS}$ (e s$^{-1}$) into TESS magnitude $M_{\rm TESS}$ (mag) following the TESS instrument handbook\footnote{<https://archive.stsci.edu/files/live/sites/mast/files/home/missions-and-data/active-missions/tess/$\_$documents/TESS$\_$Instrument$\_$Handbook$\_$v0.1.pdf>} (equation \ref{eq:f2mag}), and then adjusted to the $r$-band observations by the Lesedi telescope (+5.525 mag).

\begin{equation}
    \label{eq:f2mag}
    F_{\rm TESS} = 10^{\left[\left(- M_{\rm TESS}+20.44\right)/2.5\right]}
\end{equation}

\end{document}